\documentclass[11pt]{article}

\usepackage[T1]{fontenc}
\usepackage{lmodern}

\usepackage{amsmath, amssymb, amsthm}
\usepackage{graphicx}
\usepackage{hyperref}
\usepackage{geometry}
\usepackage{booktabs}
\usepackage{setspace}
\usepackage{float}
\usepackage{caption}
\usepackage{subcaption}
\usepackage{microtype}

\title{\textbf{Market Reactions to Material Cybersecurity Incidents}}

\author{%
  Maxwell Block\\[0.3em]
  Johns Hopkins University\\
  \texttt{mblock13@jh.edu}%
}

\date{November 2025}

\begin{document}
\maketitle

\section{Introduction}
Cybersecurity incidents are a growing concern in regulatory and financial contexts. The U.S. Securities and Exchange Commission (SEC) mandated a new disclosure requirement under Item 1.05 of Form 8-K in 2023, which requires public companies to report material cybersecurity incidents. Material is defined by the Supreme Court as ``substantial likelihood that the disclosure of the omitted fact would have been viewed by the reasonable investor as having significantly altered the `total mix' of information made available'' (TSC Industries, Inc. v. Northway, Inc.). In the context of cybersecurity, an incident is material if a reasonable investor would view it as important enough to meaningfully change their understanding of the company. Cybersecurity incidents can often be newsworthy, create operational disruptions, or have many other impacts on a company. By analyzing SEC data, the market reaction to cybersecurity disclosures can be investigated, focusing on how stock prices behave following the filing of a material cybersecurity incident.

The analysis seeks to determine whether disclosures of material cybersecurity incidents are associated with negative short-term stock price movements, and to what extent these reactions differ across companies. In doing so, the study aims to advance understanding of market trends in response to cybersecurity risks and to inform the development of empirically grounded investment strategies. Furthermore, the study examines whether specific company characteristics, such as market capitalization, industry sector, and beta (market sensitivity), moderate the impact of cybersecurity disclosures on stock price reactions.

\section{Data}
\begin{figure}[htbp]
    \centering
    \includegraphics[width=0.85\textwidth]{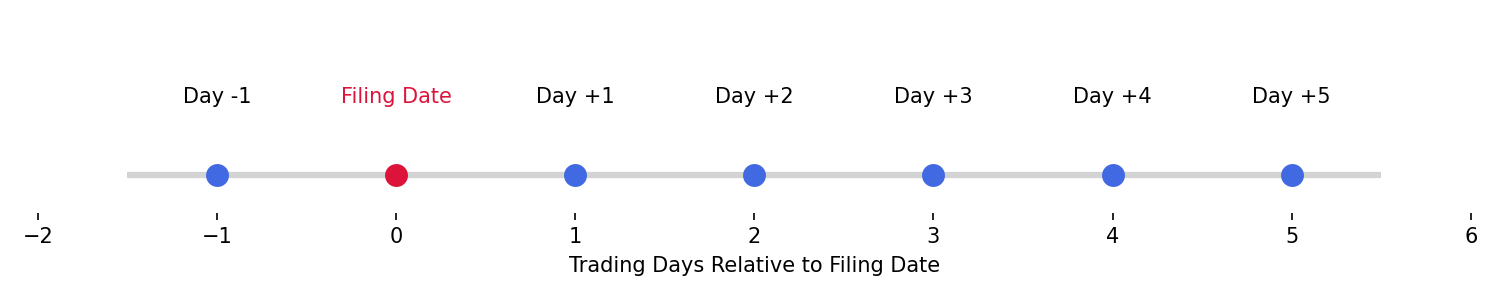}
    \caption{Visualize seven-day trading window used to measure market reaction.}
    \label{fig:eventwindow}
\end{figure}

The SEC publishes reports through their EDGAR database\footnote{\url{https://www.sec.gov/submit-filings/about-edgar}} including all public company Form 8-K filings that include Item 1.05 (Material Cybersecurity Incidents). Through a Python library\footnote{\url{https://pypi.org/project/sec-api/\#10-k10-q8-k-section-extractor-api}} information about each 8-K filing with Item 1.05 was gathered programmatically with extensive data. For each report, data related to company identifiable information such as ticker and CIK was retained, along with the filing timestamp. To measure market reaction, the daily close stock prices over a seven day window was extracted for the reporting company based on the filing timestamp day (see Figure~\ref{fig:eventwindow}). The previous market open day before the 8-K filing was made through five market days after the filing date are analyzed. If the filing date occurred on a non-trading day then the filing date is defined as the first trading day after the filing date.

For each cybersecurity event (8-K filing with Item 1.05) the market data for the stock price on this seven day window was gathered from a Python library\footnote{\url{https://pypi.org/project/yfinance/}} using Yahoo's public API. Company information such as the industry sector, current market capitalization, beta (market sensitivity), and other fields were gathered alongside stock price data. The final dataset as of November 12th, 2025 contains a total of 54 8-K reports with Item 1.05 that have been submitted to the SEC corresponding to 36 unique companies. No duplicate reports were included but there were instances in which a company experienced more than one incident. This was treated as separate events in the data as it was recorded through distinct filings with the SEC on different dates. Across the 36 companies, most experienced only one cybersecurity incident disclosure, with the distribution showing 21 companies filing a single report, 12 companies filing two reports, and 3 companies filing three reports. The companies with the highest number of incidents (three reports each) were Key Tronic Corp (KTCC), UnitedHealth Group Inc. (UNH), and First American Financial Corp (FAF).

\begin{figure}[H]
    \centering
    \includegraphics[width=0.85\textwidth]{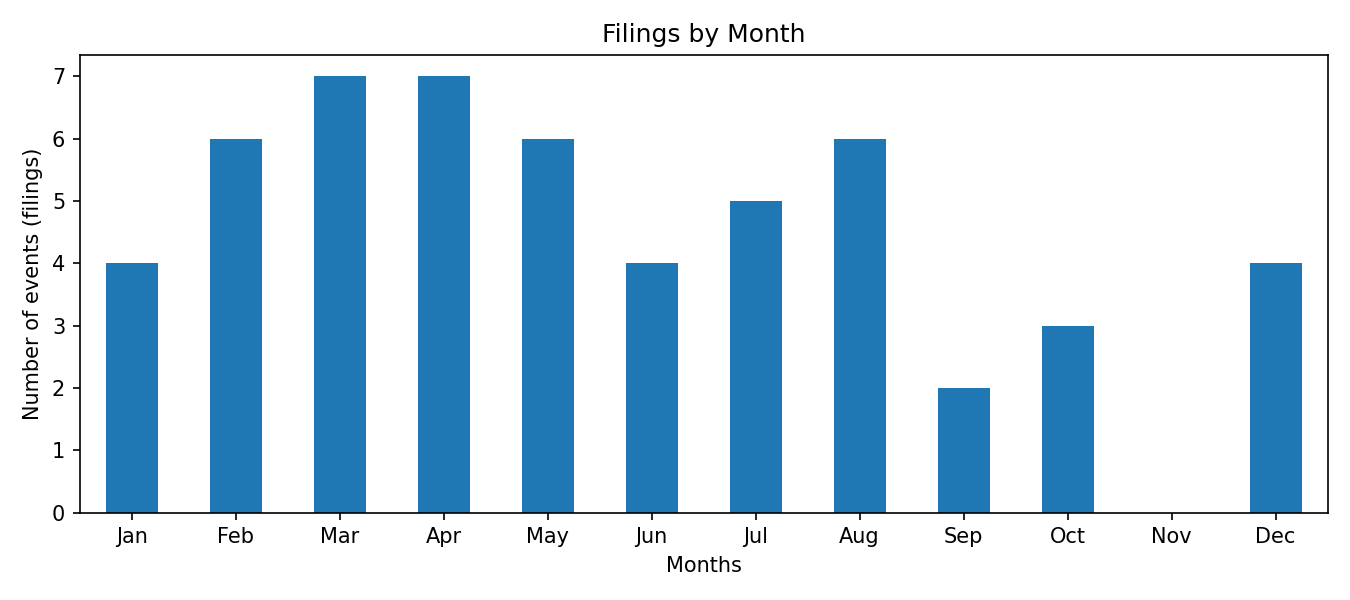}
    \caption{Frequency of material cybersecurity incident disclosures by month.}
    \label{fig:monthly}
\end{figure}

Figure~\ref{fig:monthly} shows the frequency of material cybersecurity incident disclosures by month. Examining this distribution helps determine whether filings cluster at particular times of year (e.g., fiscal reporting periods, post audit cycles, etc.). Identifying these patterns provide context for understanding seasonal trends.

Company data was gathered for all 54 filings, where each was categorized by sector, market capitalization, and beta. For market capitalization, companies were grouped into four predefined size thresholds: Small ($<\$2$ billion), Mid (\$2--10 billion), Large (\$10--200 billion), and Mega ($\ge \text{\$}200$ billion). This allowed small companies (n=24) to be compared with all larger companies (Mid, Large, and Mega) combined (n=29). Sector classification was taken from company metadata across the sample. Technology companies (n=13) were compared against all non-technology companies (n=41). The beta distribution contained one extreme outlier ($\beta = 3618.411$), while all other companies were near 1.0, making the median (1.035) a better measure of central tendency. Therefore, beta groups were formed using a median split rather than the mean. The categorizations provided a framework for analysis conducted in the hypothesis tests. Observations with missing price data, invalid market capitalization values, missing beta data, or missing sector classifications were excluded from analyses requiring those variables.

\section{Methods}
This study evaluates how stock prices respond to material cybersecurity incident disclosures filed under Item 1.05 of Form 8-K. For each disclosure, daily closing prices were collected over a seven-day window. To ensure comparability across companies with different price levels, all market reactions were measured as percentage returns, defined as the change from the closing price on the trading day before the filing to five trading days after.

The first hypothesis tested whether companies who filed a material cybersecurity incident experience a negative short-term reaction to their stock price. The null hypothesis specifies that the mean event-window return equals zero, while the alternative specifies that the mean return is negative. A one-sample, left-tailed $t$-test was used to evaluate this question:
\[
\begin{aligned}
H_{0}:~&\mu = 0,\\
H_{1}:~&\mu < 0.
\end{aligned}
\]

After establishing the overall stock price behavior, the remaining hypotheses examined whether specific company characteristics were associated with stronger negative reactions. Because each comparison involved two independent groups and unequal variances were possible, all subgroup analyses employed left-tailed Welch two-sample $t$-tests.

\subsection*{Sector Effect}
The first comparison evaluated a potential sector effect, testing whether technology companies exhibited more negative returns than other sectors:
\[
\begin{aligned}
H_{0,2}:~&\text{The average price change does not vary by sector},\\
H_{1,2}:~&\text{Technology companies exhibit larger negative reactions than other sectors}.
\end{aligned}
\]

\subsection*{Size Effect}
The second comparison evaluated a size effect, comparing small companies ($<\text{\$}2$B) with all larger companies combined:
\[
\begin{aligned}
H_{0,3}:~&\text{Price reaction is independent of market capitalization size},\\
H_{1,3}:~&\text{Smaller companies experience stronger negative reactions}.
\end{aligned}
\]

\subsection*{Beta (Market Sensitivity) Effect}
The final test examined a risk effect, dividing companies into high and low beta groups using the sample median beta as the cutoff:
\[
\begin{aligned}
H_{0,4}:~&\text{Price reaction is independent of company beta},\\
H_{1,4}:~&\text{Higher beta companies exhibit larger price drops following disclosures}.
\end{aligned}
\]

These tests provided a framework for assessing whether sector, market capitalization, or beta (market sensitivity) predicts a stronger drop in stock price after disclosing a material cybersecurity incident.

\section{Findings}

\subsection*{Overall Market Reaction}
\begin{figure}[htbp]
    \centering
    \includegraphics[width=0.8\textwidth]{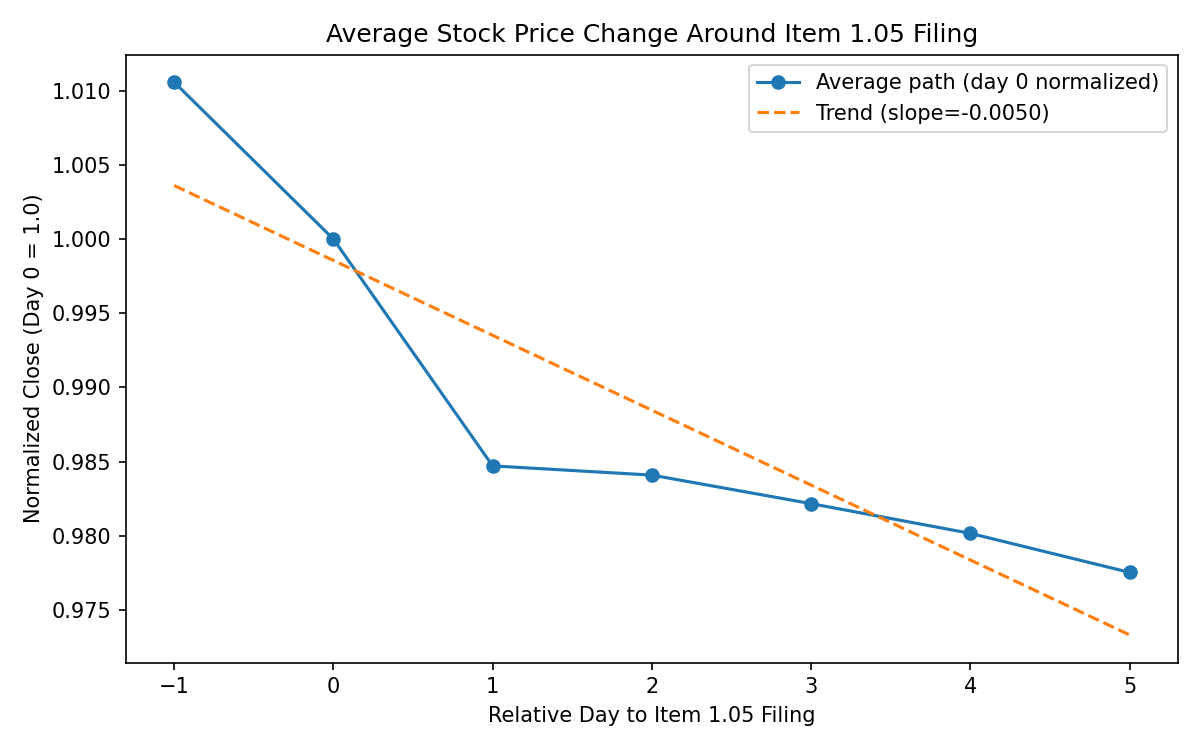}
    \caption{Normalized stock price trend over the seven-day event window.}
    \label{fig:trend}
\end{figure}

Across all 54 incidents, the average percentage stock return in the event window was $-3.07\%$, with a standard deviation of $9.36\%$. A one-sample $t$-test indicated that this decline was statistically significant,
\[
t = -2.41,\qquad p = 0.0098,
\]
providing evidence of a negative short-term stock price reaction following Item~1.05 material cybersecurity disclosures. The normalized price trend over the seven-day window is shown in Figure~\ref{fig:trend} and displays a generally downward movement consistent with the estimated negative mean return.

\subsection*{Sector Effect}
Technology companies did not exhibit materially different price reactions compared to companies in other sectors. The mean event-window return for technology firms was $-2.88\%$, compared with $-3.25\%$ for non-technology firms. A Welch two-sample $t$-test showed that this difference was not statistically significant,
\[
t = 0.162,\qquad p = 0.564.
\]
Because the $p$-value exceeds conventional significance thresholds, there is no evidence that technology companies experienced larger negative returns than other sectors following Item~1.05 disclosures.

\subsection*{Size Effect}
\begin{figure}[htbp]
    \centering
    \includegraphics[width=0.8\textwidth]{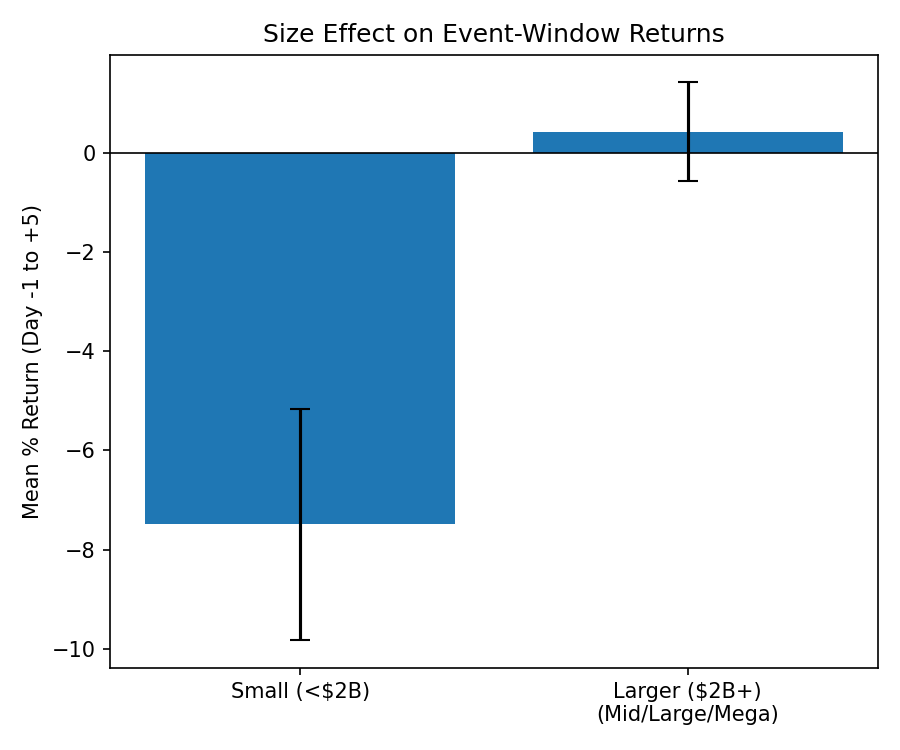}
    \caption{Comparison of event-window returns by company size category.}
    \label{fig:size}
\end{figure}

There is strong evidence that smaller companies experienced significantly larger price declines following Item~1.05 disclosures. Small firms (market capitalization $< \text{\$}2$~billion) had an average event-window return of $-7.49\%$, compared with $+0.43\%$ for larger firms (Figure~\ref{fig:size}). A Welch two-sample $t$-test confirmed that this difference is statistically significant,
\[
t = -3.13,\qquad p = 0.0019.
\]
Because the $p$-value is well below standard significance thresholds, we reject the null hypothesis and conclude that smaller companies suffered more negative short-term market reactions than larger companies.

\subsection*{Beta (Market Sensitivity) Effect}
There is no evidence that higher-beta companies experienced larger price declines following Item~1.05 disclosures. Companies with below-median beta had an average event-window return of $-4.14\%$, while higher-beta companies averaged $-2.05\%$. A Welch two-sample $t$-test showed that the difference was not statistically significant,
\[
t = 0.835,\qquad p = 0.796,
\]
indicating that higher-beta companies did not experience more negative reactions.

\section{Discussion}
The results of this study show that material cybersecurity incident disclosures are generally followed by short-term stock price declines. This is consistent with the expectation as cybersecurity incidents create risk and negative sentiment among investors. The strong size (company market cap) effect suggests that smaller companies are more vulnerable to these reactions, potentially because they have fewer resources and weaker resilience compared to larger companies. These findings show the importance of cybersecurity readiness for smaller companies. The data shows seasonality in disclosure timing, with fewer filings in November and December. Although this study cannot determine causality, this pattern reflects slower operational activity during holiday periods in American companies. It is important companies stay vigilant, especially during holiday seasons. ``The Federal Bureau of Investigation (FBI) and the Cybersecurity and Infrastructure Security Agency (CISA) have observed an increase in highly impactful ransomware attacks occurring on holidays and weekends'' (National Coordinator for Critical Infrastructure Security and Resilience).

While company market capitalization showed a difference in company short term returns after incident disclosures, other factors such as beta and sector did not show significant differences. These findings suggest opportunities for further research. For example, studies could be done to examine whether longer event windows reveal price recovery patterns, or whether reactions differ based on the nature of the incident reported, and if additional company characteristics help explain variation in outcomes.

These results also suggest potential implications for investment opportunities. Because smaller companies exhibited significantly larger short-term declines following cybersecurity disclosures, this event could create predictable outcomes which investors could attempt to exploit through short selling or derivatives strategies. While this study does not evaluate trading profitability, the magnitude of the size effect indicates that Item 1.05 disclosures may generate further financial research.

\section*{Acknowledgements}
Special thanks to Dr.~Sue-Jane Wang for providing valuable information related to this analysis. The textbook \textit{An Introduction to Mathematical Statistics and Its Applications} by Richard J.~Larsen and Morris L.~Marx served as an essential reference for the formulas, statistical inference techniques, and analytical framework applied in this study. I acknowledge the SEC's public \textsc{EDGAR} database, and the software libraries \texttt{sec-api} and \texttt{yfinance}, which enabled automated retrieval and analysis of disclosure and market data. I also acknowledge that this research was made possible through personal funding, which covered all associated costs and resources required for data acquisition, computational tools, and analysis.

\section*{References}

Larsen, Richard J., and Morris L. Marx. \textit{An Introduction to Mathematical Statistics and Its Applications}. Pearson, 2018.

National Coordinator for Critical Infrastructure Security and Resilience. Ransomware Awareness for Holidays and Weekends. CISA, 10 February 2022. \url{https://www.cisa.gov/news-events/cybersecurity-advisories/aa21-243a}. Accessed 26 November 2025.

\textit{TSC Industries, Inc. v. Northway, Inc.} 426 U.S. 438 (1976). Cornell Law School, U.S. Supreme Court. \url{https://www.law.cornell.edu/supremecourt/text/485/224}. Accessed 26 November 2025.

U.S. Securities and Exchange Commission. SEC Adopts Rules on Cybersecurity Risk Management, Strategy, Governance, and Incident Disclosure by Public Companies. SEC.gov, 26 July 2023. \url{https://www.sec.gov/newsroom/press-releases/2023-139}. Accessed 12 November 2025.

\end{document}